\documentclass[eqsecnum,preprint,prd,aps,nofootinbib]{revtex4}
\usepackage{amsmath}
\usepackage{graphics}
\def\cstok#1{\leavevmode\thinspace\hbox{\vrule\vtop{\vbox{\hrule\kern1pt
\hbox{\vphantom{\tt/}\thinspace{\tt#1}\thinspace}}
\kern1pt\hrule}\vrule}\thinspace}

\newcommand{\be}{ \begin{equation} }
\newcommand{\ee}{ \end{equation} }
\newcommand{\beqa}{ \begin{eqnarray} }
\newcommand{\eeqa}{ \end{eqnarray} }

\begin{document}
\begin{center}
\bibliographystyle{article}
{\Large \textsc{Energy-momentum tensor for a Casimir apparatus
in a weak gravitational field}}
\end{center}
\vspace{0.4cm}
\date{\today}
\author{Giuseppe Bimonte, \thanks{electronic address:
giuseppe.bimonte@na.infn.it} Enrico Calloni,
\thanks{electronic address: enrico.calloni@na.infn.it}
Giampiero Esposito, \thanks{electronic address:
giampiero.esposito@na.infn.it} and Luigi Rosa
\thanks{electronic address: luigi.rosa@na.infn.it}}
\affiliation{Dipartimento di Scienze Fisiche, Complesso Universitario
di Monte S. Angelo, Via Cintia, Edificio N', 80126 Napoli, Italy \\
INFN, Sezione di Napoli, Complesso Universitario di Monte S. Angelo,
Via Cintia, Edificio N', 80126 Napoli, Italy}

\begin{abstract}
The influence of the gravity acceleration on the regularized energy-momentum
tensor of the quantized electromagnetic field between
two plane parallel conducting plates is
derived. We use Fermi coordinates and work to first order in the
constant acceleration parameter. A perturbative expansion, to this
order, of the Green functions involved and of the energy-momentum
tensor is derived by means of the covariant geodesic point
splitting procedure. In correspondence to the Green functions
satisfying mixed and gauge-invariant boundary conditions, and Ward identities,
the energy-momentum tensor is covariantly conserved and satisfies the
expected relation between gauge-breaking and ghost parts. 
A more systematic derivation is therefore obtained of
the theoretical prediction according to which the Casimir device in a weak
gravitational field will experience a tiny push in the upwards direction.
\end{abstract}
\maketitle

\section{Introduction}

The use of Green-function methods in curved spacetime, and the
theoretical analysis of the Casimir effect, are two relevant branches
of modern quantum field theory. In the former case, some
progress along the years can be outlined as follows.
\vskip 0.3cm
\noindent
(i) In Ref. \cite{Chri76}, the covariant geodesic point separation is applied
to evaluate the vacuum expectation value of the energy-momentum tensor
$T_{\mu \nu}$ for a massive scalar field in an arbitrary gravitational
field. In Ref. \cite{Chri78}, the same author performs regularization,
renormalization and covariant geodesic point separation for spin
$0,1/2$ and $1$ fields, massive or massless, on an arbitrary curved
background; he also finds which terms in the vacuum expectation value
of $T_{\mu \nu}$ vanish within this framework.
\vskip 0.3cm
\noindent
(ii) In Ref. \cite{Bunc79}, the authors obtain a momentum-space representation
of the Feynman propagator $G(x,x')$ for scalar and spin-${1\over 2}$ fields
propagating in arbitrary curved spacetimes. Their construction uses
Riemann normal coordinates with origin at the point $x'$ and is therefore
only valid for points $x$ lying in a normal neighbourhood of $x'$. The
resulting momentum-space representation is equivalent to the
Schwinger--DeWitt proper-time representation.
\vskip 0.3cm
\noindent
(iii) In Ref. \cite{Deut79}, electromagnetic and scalar fields are quantized
in the region near an arbitrary smooth boundary. The authors find that
the components of $\langle T_{\mu \nu}(x) \rangle_{\rm ren}$ generically
diverge in a nonintegrable manner as $x$ approaches the boundary of the
manifold from the interior. They therefore conclude that perfect conductor
boundary conditions are pathological, in that the distribution of
energy-momentum that they would entail if they were actually to obtain
for arbitrarily high frequencies and short wavelengths would be such as
to produce an infinite {\it physically observable} gravitational field.
\vskip 0.3cm
\noindent
(iv) In Ref. \cite{Brow86}, the authors derive the symmetric Hadamard
representation for scalar and photon Green functions, and use
these representations to give a simple definition for their associated
renormalized energy-momentum tensors.
\vskip 0.3cm
\noindent
(v) In Ref. \cite{Bimo04}, the full asymptotic expansion of the Feynman
photon Green function at small values of the world function, as well as
its explicit dependence on the gauge parameter, are obtained without
adding by hand a mass term to the DeWitt--Faddeev--Popov Lagrangian.
Coincidence limits of second covariant derivatives of the associated
Hadamard function are also evaluated.

On the other hand, an important property of quantum electrodynamics is
that suitable differences of zero-point energies of the quantized
electromagnetic field can be made finite and produce measurable effects
such as the tiny attractive force among perfectly conducting parallel
plates known as the Casimir effect \cite{Bord01}. This is a remarkable
quantum mechanical effect that makes itself manifest on a macroscopic
scale. For perfect reflectors and metals the Casimir force can be
attractive or repulsive, depending on the geometry of the cavity, whereas
for dielectrics in the weak-reflector approximation it is always attractive,
independently of the geometry \cite{Bart01}. The Casimir effect can be
studied within the framework of boundary effects in quantum field theory,
combined with zeta-function regularization or Green-function methods,
or in more physical terms, i.e. on considering van der Waals forces
\cite{Kamp68} or scattering problems \cite{Grah02}. Casimir energies
are also relevant in the attempt of building a quantum theory of gravity
and of the universe \cite{Davi05}.

For these reasons, in Ref. \cite{Call02} we evaluated the force
produced by a weak gravitational field on a rigid Casimir cavity.
Interestingly, the resulting force was found to have opposite
direction with respect to the gravitational acceleration;
moreover, we found that the current experimental sensitivity of
small force macroscopic detectors would make it possible, at least
in principle, to measure such an effect \cite{Call02}. More
precisely, the gravitational force on the Casimir cavity might be
measured provided one were able to use rigid cavities and  find an
efficient force modulation method \cite{Call02}. Rigid cavities,
composed by metal layers separated by a dielectric layer, make it
possible to reach separations as small as 5$\div$10 nm and allow
to build multi-cavity structures, made by a sequence of such
alternate layers. If an efficient modulation method could be
found, it would be possible  to achieve a modulated force of order
$10^{-14}$ N in the earth's gravitational field. The measure of
such a force, already possible with current small-force detectors
on macroscopic bodies, might open the way to the first test of the
gravitational field influence on vacuum energy \cite{Call02}. In
Ref. \cite{Call02}, calculations were based on simple assumptions
and the result can be viewed as a reasonable ``{\em first order}''
generalization of $T_{\mu\nu}$ from Minkowski to curved
space-time. The present paper is devoted to a deeper understanding
and to more systematic calculations of the interaction of a weak
gravitational field with a Casimir cavity. To first order in our
approximation the former value of the force exerted by the field
on the cavity is recovered. 

We consider a plane-parallel Casimir cavity, made of
ideal metallic plates, at rest in the gravitational field
of the earth, with its plates lying in a horizontal plane.
We evaluate the influence of the gravity acceleration
$g$ on the Casimir cavity but neglect any variation
of the gravity acceleration across the cavity, and therefore
we do not consider the influence of tidal forces.
The separation $a$ between the plates is taken to be much
smaller than the extension of the plates, so that edge
effects can be neglected. We obtain a perturbative
expansion of the energy-momentum tensor of the electromagnetic field
inside the cavity, in terms of the small parameter
$\epsilon \equiv 2 ga/c^2$, to first order in $\epsilon$. For this purpose,
we use a Fermi \cite{Misn73, Marz94} coordinates system
$(t,x,y,z)$ rigidly connected to the cavity.
The construction of these coordinates
involves only invariant quantities such as the observer's proper time,
geodesic distances from the world-line, and components of tensors
with respect to a tetrad \cite{Marz94}. This feature makes it possible
to obtain a clear
identification of the various terms occurring in the metric.
In our analysis we adopt the covariant point-splitting procedure
\cite{Dewi75, Chri76} to compute the perturbative expansion
of the relevant Green functions. Gauge invariance
plays a crucial role and we check it up to first order by means of
the Ward identity. We also evaluate the Casimir energy and pressure, and
in this way we obtain a sound derivation of the result in Ref. \cite{Call02},
according to which the Casimir device in a weak gravitational field
will experience a tiny push in the
upwards direction. Use is here made of mixed
boundary conditions on the potential plus Dirichlet conditions
on ghost fields. Concluding remarks are presented in Sec. VI, while
relevant details are given in the Appendices.

\section{Spin-1 and spacetime formalism}

The classical action functional for the Maxwell potential $A_{\mu}(x)$
reads
\begin{equation}
S[A_{\mu}]=-{1\over 4}\int F_{\mu \nu}F^{\mu \nu}
\sqrt{-g}d^{4}x,
\label{(2.1)}
\end{equation}
where the field strength $F_{\mu \nu} \equiv \nabla_{\mu}A_{\nu}
-\nabla_{\nu}A_{\mu}=A_{\nu ; \mu}-A_{\mu ; \nu}$. By virtue of the gauge
invariance of the action, the differential operator
$
{\delta^{2}S \over \delta A_{\mu} \delta A_{\nu'}}
$
is singular (see Appendix A).
To take care of this problem one should add a gauge-breaking
term, which leads to a nonsingular wave operator
${\widehat U}^{\alpha \beta'}$ on the potential.
In the Lorenz (this is not Lorentz!) gauge \cite{Lore67},
and with the Feynman choice for the gauge parameter (see Appendix A),
this is $-{1\over 2}(\nabla^{\mu}A_{\mu})^{2}$. Last, a ghost term
$\chi^{;\alpha}\psi_{;\alpha}$ is necessary, where $\chi$ and $\psi$
are independent ghost fields \cite{Dewi84}. The full action is therefore
\begin{equation}
S[A_{\mu},\chi,\psi]=\int \left[-{1\over 4}F_{\mu \nu}F^{\mu \nu}
-{1\over 2}(\nabla^{\mu}A_{\mu})^{2}+\chi^{; \alpha}\psi_{; \alpha}
\right]\sqrt{-g}d^{4}x,
\label{(2.2)}
\end{equation}
with field equations \cite{Chri78}
\begin{equation}
\int {\widehat U}^{\alpha \beta'}A_{\beta'}d^{4}x'=0,
\label{(2.3)}
\end{equation}
\begin{equation}
\int {\widehat U}(x,x')\psi(x')d^{4}x'=0,
\label{(2.4)}
\end{equation}
having defined (here $\delta^{\alpha \beta'} \equiv g^{\alpha \beta}
\delta(x,x')$)
\begin{equation}
{\widehat U}^{\alpha \beta'} \equiv
{\delta^{2}S \over \delta A_{\alpha} \delta A_{\beta'}}
=\sqrt{-g}\left(
\delta_{\; \; \; \; \; ;\rho}^{\alpha \beta' \; \; \; \rho}
-R_{\; \rho}^{\alpha} \delta^{\rho \beta'}\right),
\label{(2.5)}
\end{equation}
\begin{equation}
{\widehat U}(x,x') \equiv {\delta^{2}S \over \delta \chi \delta \psi'}
=-\sqrt{-g}\delta_{;\rho}^{\; \; \; \rho}(x,x').
\label{(2.6)}
\end{equation}
The energy-momentum tensor is obtained from functional differentiation
of the full action (2.2) according to
\begin{equation}
T^{\mu \nu} \equiv {2\over \sqrt{-g}}{\delta S \over \delta g_{\mu \nu}},
\label{(2.7)}
\end{equation}
and turns out to be the sum of three contributions $T_{A},T_{B},T_{\rm gh}$
resulting from Maxwell action, gauge-breaking term and ghost action,
respectively \cite{Chri78} (no mass term occurs
since we do not add by hand any mass for photons or ghost fields, unlike
Ref. \cite{Chri78}). This leads eventually to the following vacuum
expectation value:
\begin{equation}
\langle T^{\mu \nu} \rangle = \langle T_{A}^{\mu \nu} \rangle
+\langle T_{B}^{\mu \nu}\rangle+\langle T_{\rm gh}^{\mu \nu}\rangle,
\label{(2.8)}
\end{equation}
where, on defining the photon and ghost Hadamard functions, respectively,
as
\begin{equation}
H_{\mu \nu}(x,x') \equiv \langle [A_{\mu}(x),A_{\nu}(x')]_{+} \rangle
\equiv H_{\mu \nu'},
\label{(2.9)}
\end{equation}
\begin{equation}
H(x,x') \equiv \langle [\chi(x),\psi(x')]_{+}\rangle,
\label{(2.10)}
\end{equation}
jointly with
\begin{eqnarray}
\langle F_{\rho\alpha}F_{\tau\beta}\rangle &=& \lim_{x'\rightarrow
x} \frac{1}{4}\left[H_{\alpha\beta';\rho\tau'} +
H_{\beta\alpha';\tau\rho'}-H_{\alpha\tau';\rho\beta'}-
H_{\tau\alpha';\beta\rho'}- H_{\rho\beta';\alpha\tau'}
\right. \nonumber \\
& & \left.-
H_{\beta\rho';\tau\alpha'}+H_{\rho\tau';\alpha\beta'}+
H_{\tau\rho';\beta\alpha'}\right],
\label{(2.11)}
\end{eqnarray}
one has
\begin{equation}
\langle T_{A}^{\mu\nu} \rangle  = \lim_{x'\rightarrow x}
\left[\left(g^{\mu\rho}g^{\nu\tau}
-\frac{1}{4}g^{\mu\nu}g^{\tau\rho}
\right) g^{\alpha\beta} \langle
F_{\rho\alpha}F_{\tau\beta}\rangle \right],
\label{(2.12)}
\end{equation}
\begin{equation}
\langle T_{B}^{\mu\nu} \rangle = \lim_{x'\rightarrow x} \left[
-\frac{1}{4}g^{\alpha\beta}\left(
g^{\mu\rho}g^{\nu\tau}+g^{\mu\tau}g^{\nu\rho}-g^{\mu\nu}g^{\tau\rho}\right)
\left(H_{\beta\tau';\alpha\rho}+H_{\tau\beta';\rho\alpha'}\right)\right],
\label{(2.13)}
\end{equation}
\begin{equation}
\langle T_{\rm gh}^{\mu\nu}\rangle = \lim_{x'\rightarrow x} \left[
-\frac{1}{4}\left(
g^{\mu\alpha}g^{\nu\beta}+g^{\mu\beta}g^{\nu\alpha}
-g^{\mu\nu}g^{\alpha\beta}\right)
\left(H_{;\alpha\beta'}+ H_{;\beta\alpha'} \right)\right].
\end{equation}
Following Christensen \cite{Chri78}, we separate points symmetrically, i.e.
\begin{equation}
[A_{\alpha;\rho},A_{\beta;\tau}]_{+} =\lim_{x'\rightarrow x}
\frac{1}{2}\left\{ [A_{\alpha';\rho'},A_{\beta;\tau}]_{+}
+[A_{\alpha;\rho},A_{\beta';\tau'}]_{+} \right\}.
\label{(2.15)}
\end{equation}
In the applications considered in Sec. IV, the coincidence limit will
be taken in two separate steps: first with respect to time and the
coordinates spanning the plates, and eventually with respect to the
$z$-coordinate (see Appendix B for details).

In the implementation of Eq. (2.15) one needs the geodesic parallel
displacement bivector $g_{\; \nu'}^{\mu}$ (in general, bi-tensors
behave as a tensor both at $x$ and at $x'$)
which effects parallel displacement
of vectors along the geodesic from $x'$ to $x$. In general, it is
defined by the differential equations
\begin{equation}
\sigma^{; \rho}g_{\; \nu' ; \rho}^{\mu}
=\sigma^{; \tau'} g_{\; \nu' ; \tau'}^{\mu}=0,
\label{(2.16)}
\end{equation}
$\sigma(x,x')$ being the Ruse--Synge world function
\cite{RuSy31}, equal to half the geodesic
distance between $x$ and $x'$, jointly with the coincidence limit
(boundary condition)
\begin{equation}
\lim_{x' \rightarrow x}g_{\; \nu'}^{\mu} \equiv
\left[g_{\; \nu'}^{\mu}\right]=\delta_{\; \nu}^{\mu}.
\label{(2.17)}
\end{equation}
The bivector $g_{\; \nu'}^{\mu}$, when acting on a vector $B^{\nu'}$
at $x'$, gives the vector ${\overline B}^{\mu}$, which is obtained by
parallel transport of $B^{\nu'}$ to $x$ along the geodesic connecting
$x$ and $x'$, i.e.
\begin{equation}
{\overline B}^{\mu}=g_{\; \nu'}^{\mu}B^{\nu'}.
\label{(2.18)}
\end{equation}

For the reasons described in the Introduction, we use Fermi coordinates.
With our choice, the $z$-axis coincides with the vertical upwards
direction, while the $(x,y)$ coordinates span the plates, whose equations
are $z=0$ and $z=a$, respectively.
The resulting line element for a nonrotating system
is therefore \cite{Misn73}
\begin{equation}
ds^{2}= -c^{2}\left(1+\epsilon {z\over a}\right)dt^{2}
+dx^{2}+dy^{2}+dz^{2}
+{\rm O}(\mid x \mid^{2})
= \eta_{\mu \nu}dx^{\mu}dx^{\nu}-\epsilon {z\over a} c^{2}dt^{2},
\label{(2.19)}
\end{equation}
where the perturbation parameter $\epsilon \equiv {2ga / c^{2}}$,
while $\eta_{\mu \nu}$ is the flat Minkowski metric ${\rm
diag}(-1,1,1,1)$.

\section{Green Functions}

For any field theory, once that the invertible differential operator
$U_{ij}$ in the functional integral is given
(see Eq. (A4) in Appendix A), the corresponding Green
functions satisfy the condition
\begin{equation}
U_{ij}G^{jk}=-\delta_{i}^{\; k},
\label{(3.1)}
\end{equation}
and are boundary values of holomorphic functions. The choice of boundary
conditions will determine whether we deal with advanced Green functions
$G^{+jk}$, for which the integration contour passes below the poles of the
integrand on the real axis, or retarded Green functions $G^{-jk}$, for
which the contour passes instead above all poles on the real axis, or yet
other types of Green functions. Among these, a key role is played by the
Feynman Green function $G_{F}^{jk}$, obtained by choosing a contour that
passes below the poles of the integrand that lie on the negative real axis
and above the poles on the positive real axis. If one further defines
the Green function
\cite{Dewi84}
\begin{equation}
{\overline G}^{jk} \equiv {1\over 2}(G^{+jk}+G^{-jk}),
\label{(3.2)}
\end{equation}
one finds in stationary backgrounds (for which the metric is independent of
the time coordinate, so that there exists a timelike Killing vector field)
that the Feynman Green function has a real part equal to
${\overline G}^{jk}$, and an imaginary part equal to the Hadamard function
$H^{jk}$, i.e.
\begin{equation}
H^{jk}(x,x') \equiv -2i \Bigr[G_{F}^{jk}(x,x')
-{\overline G}^{jk}(x,x')\Bigr].
\label{(3.3)}
\end{equation}
This relation can be retained as a definition when the background is
nonstationary; in such a case, however, $H^{jk}(x,x')$ is generally
no longer real \cite{Dewi84}.

In particular, the photon Green function $G_{\lambda \nu'}$ in a curved
spacetime with metric $g_{\mu \nu}$ solves the equation \cite{Bimo04}
\begin{equation}
\sqrt{-g}P^{\; \lambda}_\mu(x) G_{\lambda\nu'}=g_{\mu\nu}
\delta(x,x').
\label{(3.4)}
\end{equation}
The wave operator $P_{\mu}^{\; \lambda}$ results from the gauge-fixed
action (2.2) with Lorenz gauge-fixing functional
$\Phi_{L}(A) \equiv \nabla^{\mu}A_{\mu}$, and having set to $1$ the
gauge parameter of the general theory, so that (cf. Eq. (2.5))
\begin{equation}
P^{\; \lambda}_\mu(x)=-\delta^{\; \lambda}_\mu \cstok{ }_x
+R^{\; \lambda}_\mu(x),
\label{(3.5)}
\end{equation}
where $\cstok{ }_x \equiv g^{\alpha\beta}
\nabla_{\alpha}\nabla_{\beta}(x)$.
Since we need the action of the gauge-field operator
$P^{\; \lambda}_\mu(x)$ on the photon Green functions,
it is worth noticing that
\begin{equation}
D_{\beta\lambda\nu'}  \equiv  \nabla_\beta
G_{\lambda\nu'} = \partial_\beta
G_{\lambda\nu'}-\Gamma^\mu_{\beta\lambda}G_{\mu\nu'},
\label{(3.6)}
\end{equation}
\begin{equation}
Q_{\alpha\beta\lambda\nu'}  \equiv \nabla_\alpha \nabla_\beta
G_{\lambda\nu'} = \nabla_\alpha D_{\beta\lambda\nu'}
=\partial_\alpha D_{\beta\lambda\nu'}
-\Gamma^\mu_{\alpha\beta}D_{\mu\lambda\nu'}
-\Gamma^\mu_{\alpha\lambda}D_{\beta\mu\nu'}. 
\label{(3.7)}
\end{equation}

The Christoffel coefficients for our metric (2.19) read
\begin{equation}
\Gamma^\alpha_{\beta\gamma} =
\frac{1}{2}g^{\alpha\delta}\left( g_{\delta\beta,\gamma}+
g_{\delta\gamma,\beta}-
g_{\beta\gamma,\delta}\right)
= -\frac{1}{2}{\epsilon \over a}
\left( \eta^{\alpha 0}\delta^0_\beta
\delta^3_\gamma + \eta^{\alpha 0}\delta^0_\gamma \delta^3_\beta-
\eta^{3\alpha}\delta^0_{\gamma}\delta^0_{\beta}\right).
\label{(3.8)}
\end{equation}
Since the connection coefficients, to first order in
$\epsilon$, are constant, we realize that the Ricci curvature tensor
vanishes to this order. On expanding (this is, in general, only an
asymptotic expansion)
\begin{equation}
G_{\lambda\nu'} \sim G^{(0)}_{\lambda\nu'}+\epsilon \;
G^{(1)}_{\lambda\nu'} + {\rm O}(\epsilon^{2}), 
\label{(3.9)}
\end{equation}
we get
\begin{equation}
D_{\beta\lambda\nu'}= \partial_\beta
G_{\lambda\nu'}-\Gamma^\mu_{\beta\lambda}G_{\mu\nu'}
=\partial_\beta
G_{\lambda\nu'}-\Gamma^\mu_{\beta\lambda}G^{(0)}_{\mu\nu'},
\label{(3.10)}
\end{equation}
so that
\begin{equation}
Q_{\alpha\beta\lambda\nu'} =\partial_\alpha\partial_\beta
G_{\lambda\nu'}
-\partial_\alpha\left[\Gamma^\mu_{\beta\lambda}G^{(0)}_{\mu\nu'}\right]
-\Gamma^\mu_{\alpha\beta}\partial_\mu
G^{(0)}_{\lambda\nu'}
-\Gamma^\mu_{\alpha\lambda}\partial_\beta
G^{(0)}_{\mu\nu'},
\label{(3.11)}
\end{equation}
and finally
\begin{eqnarray}
\cstok{ }_x G_{\lambda\nu'}&=&
g^{\alpha\beta}\nabla_\alpha\nabla_\beta G_{\lambda\nu'}
=\left(\eta^{\alpha\beta}
+ \epsilon {z\over a} \delta^{\alpha}_0\delta^{\beta}_0 \right)
\nabla_\alpha\nabla_\beta\left[G^{(0)}_{\lambda\nu'}
+\epsilon
G^{(1)}_{\lambda\nu'} \right] \nonumber \\
&=&\eta^{\alpha\beta}\left[ \partial_\alpha\partial_\beta
G^{(0)}_{\lambda\nu'}+\epsilon
\partial_\alpha\partial_\beta
G^{(1)}_{\lambda\nu'}
-\Gamma^\mu_{\beta\lambda}G^{(0)}_{\mu\nu',\alpha}\right. \nonumber \\
& &\left.-\Gamma^\mu_{\alpha\beta}
G^{(0)}_{\lambda\nu',\mu}-\Gamma^\mu_{\alpha\lambda}
G^{(0)}_{\mu\nu',\beta}\right]
-\epsilon {z\over a} \delta^{\alpha}_0\delta^{\beta}_0
\partial_\alpha\partial_\beta
G^{(0)}_{\lambda\nu'}.
\label{(3.12)}
\end{eqnarray}
We therefore get, to first order in $\epsilon$,
\begin{equation}
\cstok{ }^{0} G^{(0)}_{\mu\nu'}  = J^{(0)}_{\mu\nu'},
\label{(3.13)}
\end{equation}
\begin{equation}
\cstok{ }^{0} G^{(1)}_{\mu\nu'} = J^{(1)}_{\mu\nu'},
\label{(3.14)}
\end{equation}
where
\begin{equation}
J^{(0)}_{\mu\nu'}  \equiv -\eta_{\mu\nu}\delta(x,x'),
\label{(3.15)}
\end{equation}
\begin{equation}
\epsilon J^{(1)}_{\mu\nu'}  \equiv {z\over a} \epsilon \left
(\frac{\eta_{\mu\nu}}{2}+\delta^0_{\mu}\delta^0_{\nu}\right)\delta(x,x')
+2 \eta^{\rho\sigma}\Gamma^\tau_{\sigma\mu}
G^{(0)}_{\tau\nu',\rho} +
\eta^{\rho\sigma}\Gamma^\tau_{\rho\sigma} G^{(0)}_{\mu\nu',\tau}
-{z\over a} \epsilon G^{(0)}_{\mu\nu',00}, 
\label{(3.16)}
\end{equation}
with $\cstok{ }^0\equiv\eta^{\alpha\beta}
\partial_\alpha\partial_\beta=
-\partial_0^2+\partial_x^2+\partial_y^2+\partial_z^2$.

To fix the boundary conditions we note that, on denoting by
${\vec E}_{t}$ and ${\vec H}_{n}$ the tangential and normal components
of the electric and magnetic fields, respectively,
a sufficient condition to obtain
\begin{equation}
\left . \vec{E}_{t} \right |_{S}=0,~~
\left . \vec{H}_{n} \right |_{S}=0,
\label{(3.17)}
\end{equation}
on the boundary $S$ of the device, is to impose Dirichlet
boundary conditions on $A_0(\vec{x}),A_1(\vec{x}),A_2(\vec{x})$
\cite{Espo97} at the boundary $z=0$, $z=a$. The boundary condition
on $A_3$ is determined by requiring that the gauge-fixing functional,
here chosen to be of the Lorenz type, should vanish on the boundary
(see Appendix A). This implies
\begin{equation}
\left . A^\mu_{;\mu} \right |_{S}=
0\Rightarrow \left . A^3_{;3} \right |_{S}
= \left . (g^{33}\partial_3
A_3-g^{\mu\nu}\Gamma^3_{\mu\nu}A_{3} )\right |_{S}=0 ~.
\label{(3.18)}
\end{equation}
To first order in $\epsilon$, these conditions imply the following
equations for Green functions:
\begin{equation}
\left . G^{(0)}_{\mu\nu'} \right|_{S} = 0, ~~ \mu=0,1,2,
\; \forall \nu',
\label{(3.19)}
\end{equation}
\begin{equation}
\left . \partial_3 G^{(0)}_{3\nu'} \right |_{S} = 0, ~~\forall \; \nu',
\label{(3.20)}
\end{equation}
\begin{equation}
\left . G^{(1)}_{\mu\nu'} \right |_{S} = 0,
~~ \mu =0,1,2, \; \forall \nu',
\label{(3.21)}
\end{equation}
\begin{equation}
\left . \partial_3 G^{(1)}_{3\nu'} \right |_{S}
= -\frac{1}{2a} \left . G^{(0)}_{3\nu'} \right |_{S},
 ~~\forall \; \nu',
\label{(3.22)}
\end{equation}
hence we find that the third component of the potential $A_\mu$
satisfies homogeneous Neumann boundary conditions to zeroth order in
$\epsilon$ and inhomogeneous boundary conditions to first order.

Since $J^{(0)}_{\mu\nu'} $ is diagonal, by virtue of the
homogeneity of the boundary conditions, $G^{(0)}_{\lambda\nu'}$
turns out to be diagonal. On the contrary, $J^{(1)}_{\mu\nu'}$ has two
off-diagonal contributions: $J^{(1)}_{03} $ and $J^{(1)}_{30}$, so
that $G^{(1)}_{\mu\nu'}$ is nondiagonal. Let us write
down explicitly the expressions for the various components of
$J^{(1)}_{\lambda\nu'}$, i.e.
\begin{equation}
aJ^{(1)}_{00'}  = \frac{z}{2}\delta(x,x') -z
G^{(0)}_{00',00}+ \frac{1}{2}G^{(0)}_{00',3},
\label{(3.23)}
\end{equation}
\begin{equation}
aJ^{(1)}_{03'}  = -G^{(0)}_{33',0},
\label{(3.24)}
\end{equation}
\begin{equation}
aJ^{(1)}_{11'}  =   \frac{z}{2}\delta(x,x') -z
G^{(0)}_{11',00} - \frac{1}{2}G^{(0)}_{11',3},
\label{(3.25)}
\end{equation}
\begin{equation}
aJ^{(1)}_{22'}  =  \frac{z}{2}\delta(x,x') -z
G^{(0)}_{22',00} - \frac{1}{2}G^{(0)}_{22',3},
\label{(3.26)}
\end{equation}
\begin{equation}
aJ^{(1)}_{33'}  =  \frac{z}{2}\delta(x,x') -z G^{(0)}_{33',00}
- \frac{1}{2}G^{(0)}_{33',3},
\label{(3.27)}
\end{equation}
\begin{equation}
aJ^{(1)}_{30'}  = -G^{(0)}_{00',0}.
\label{(3.28)}
\end{equation}
Now we are in a position to  evaluate, at least formally (see
below), the solutions to zeroth and first order, and we get
\begin{equation}
G^{(0)}_{\lambda\nu'} = \eta_{\lambda\nu'} \int{
\frac{d\omega d^2k }{(2\pi)^3} e^{-i\omega(t-t')+ i
{\vec k}_{\perp}\cdot({\vec x}_{\perp}-{\vec x}_{\perp}')} }
g_{D,N}(z,z'),
\label{(3.29)}
\end{equation}
having defined
\begin{equation}
g_{D}(z,z';\kappa) \equiv \frac{\sin{\kappa(a
z_<)}\sin{\kappa(a-z_>)}}{\kappa\sin{\kappa
a} },~~~~~~~0<z,z'<a,
\label{(3.30)}
\end{equation}
\begin{equation}
g_{N}(z,z';\kappa) \equiv - \frac{\cos{\kappa(a
z_<)}\cos{\kappa(a-z_>)}}{\kappa\sin{\kappa
a} },~~~~0<z,z'<a,
\label{(3.31)}
\end{equation}
where $D,N$ stand for homogeneous Dirichlet or Neumann boundary
conditions, respectively, $z_>~(z_<)$ are the larger (smaller)
between $z$ and $z'$, while ${\vec k}_{\perp}$ has components
$(k_x,k_y)$, ${\vec x}_{\perp}$ has components
$(x,y)$, $\kappa \equiv \sqrt{\omega^2-k^2}$, and
\begin{equation}
G^{(1)}_{\mu \nu'}=\int{ \frac{d\omega d^2k }{(2\pi)^3}
e^{-i\omega(t-t')+i{\vec k}_{\perp}\cdot({\vec x}_{\perp}
-{\vec x}_{\perp}')} \Phi_{\mu \nu'} },
\label{(3.32)}
\end{equation}
where the $\Phi$ components different from zero are written in Appendix B.
A scalar field satisfies the same equations of the $22$ component of
the gauge field, hence we do not write it explicitly. In the following
we will write simply $G_{\mu\nu'}$ and $G$ for the Green
function of the gauge and ghost field, respectively.

We should stress at this stage that, in general, the integrals
defining the Green functions are divergent. They are well defined
until $x\neq x'$, hence we will perform all our calculations
maintaining the points separated and only in the very end shall we
take the coincidence limit as $x' \rightarrow x$ \cite{Endo84}. We
have decided to write the divergent terms explicitly so as to bear
them in mind and remove them only in the final calculations by
hand, instead of making the subtraction at an earlier stage.

Incidentally, we note that these Green functions satisfy the
Ward identity (see Appendix A)
\begin{equation}
G^\mu_{\; \nu';\mu}+G_{;\nu'}=0,~~~
G^{\mu \; \; ;\nu'}_{\; \nu'}+G^{;\mu}=0,
\label{(3.33)}
\end{equation}
to first order in
$\epsilon$ so that, to this order, gauge invariance is explicitly preserved
(the check being simple but nontrivial).

\section{Energy-momentum tensor}

By virtue of the formulae of Sec. III we get, from the asymptotic expansion
$T_{\mu\nu'} \sim T^{(0)}_{\mu\nu'}+{\epsilon \over a} T^{(1)}_{\mu\nu'}
+{\rm O}(\epsilon^{2})$ (here
we present only the final results for brevity, while
the complete calculation is reproduced in Appendix B),
\begin{equation}
\langle T^{(0)\mu\nu'}\rangle =\frac{1 }{16\,a^4\,{\pi }^2}
\left( {\zeta}_{H}\left(4,\frac{2\,a + z - {z'}}{2\,a}\right) +
{\zeta}_{H}\left(4,\frac{z'-z}{2\,a}\right) \right)
{\rm diag}(-1,1,1,-3),
\label{(4.1)}
\end{equation}
where $\zeta_{H}$ is the Hurwitz $\zeta$-function
$\zeta_{H}(x,\beta) \equiv \sum_{n=0}^{\infty}(n+\beta)^{-x}$.
On taking the limit $z'\rightarrow z^+$ we get
\begin{equation}
\lim_{z' \to z^+} \langle T^{(0)\mu\nu'} \rangle =\left(\frac{\pi^2 }{720 a^4}
+\lim_{z' \to z^+} \frac{1}{\pi^2(z-z')^4}\right)
{\rm diag}(-1,1,1,-3),
\label{(4.2)}
\end{equation}
where the divergent term as $z' \rightarrow z$ can be removed by
subtracting the contribution of infinite space without bounding
surfaces \cite{Bord01}, and in our analysis we therefore discard
it hereafter. The renormalization of the energy-momentum tensor in
curved spacetime is usually performed by subtracting the $\langle
T_{\mu \nu} \rangle$ constructed with an Hadamard or
Schwinger--DeWitt two-point function up to the fourth adiabatic
order \cite{Chri76, Chri78}. In our problem, however, as we work
to first order in $\epsilon$, we are neglecting tidal forces and
therefore the geometry of spacetime in between the plates is flat.
Thus, we need only subtract the contribution to the energy
momentum tensor that is independent of $a$, which is the standard
subtraction in the context of the Casimir effect in flat
spacetime.

In the same way (see Appendix B) we get, to first order in $\epsilon$:
\begin{eqnarray}
\lim_{z' \to z^+} \langle T^{(1)\mu\nu'}\rangle &=&
{\rm diag}(T^{(1)00},T^{(1)11},T^{(1)22},T^{(1)33}) \nonumber \\
&+& \lim_{z' \to z^+} {\rm diag}\Bigr(-z'/\pi^{2}(z-z')^{4},0,0,0 \Bigr),
\label{(4.3)}
\end{eqnarray}
where
\begin{equation}
T^{(1)00} = -\frac{{\pi }^2}{1200\,a^3} +
\frac{{11 \pi }^2\,z}{3600\,a^4} -
\frac{\pi \,\cot (\frac{\pi \,z}{a})\,{\csc^{2} (\frac{\pi
\,z}{a})}}{60\,a^3},
\label{(4.4)}
\end{equation}
\begin{equation}
T^{(1)11} = \frac{{\pi }^2}{3600\,a^3} - \frac{{\pi}^2\,z}{1800\,a^4}
  -\frac{\pi \,\cot (\frac{\pi \,z}{a})\,{\csc^{2} (\frac{\pi
\,z}{a})}}{120\,a^3},
\label{(4.5)}
\end{equation}
\begin{equation}
 T^{(1)22} = T^{(1)11},
\label{(4.6)}
\end{equation}
\begin{equation}
T^{(1)33} = -\frac{\left( {\pi }^2\,\left( a - 2\,z \right)
\right) }{720\,a^4}.
\label{(4.7)}
\end{equation}

By virtue of the Ward identities (3.33), here checked up to first
order in $\epsilon$, the gauge-breaking part
of the energy-momentum tensor is found to be minus the ghost part, hence
we compute the second only, and the result is written
in Appendix B.

\section{Casimir energy and force}

To compute the Casimir energy we must project the energy-momentum
tensor along the unit timelike vector $u$ with covariant
components $u_\mu=(\sqrt{-g_{00}},0,0,0)$ to obtain $\rho=\langle
T^{\mu\nu}\rangle u_\mu u_\nu$, so that 
\begin{equation} 
\rho = -\frac{\pi^2 }{720a^4}+2 \frac{g}{c^2}
\left(-\frac{{\pi }^2}{1200\,a^3} +
\frac{{\pi }^2\,z}{600\,a^4}  -
\frac{\pi \,\cot (\frac{\pi \,z}{a})\,{\csc^{2} (\frac{\pi
\,z}{a})}}{60\,a^3}\right)
+{\rm O}(g^{2}), 
\label{(5.1)} 
\end{equation} 
where we have substituted $\epsilon$ by its expression in terms of $g$.
Thus, the energy stored in the Casimir device is found to be 
\be
E=\int{d^3\Sigma\sqrt{-g}\langle T^{\mu\nu}\rangle u_\mu
u_\nu}=-\frac{\hbar c \pi^2}{720} \frac{A}{a^{3}}
\left(1+\frac{1}{2} \frac{g a}{c^2} \right), 
\label{(5.2)} 
\ee
where $A$ is the area of the plates, $d^3\Sigma$ is the
three-volume element of an observer with four-velocity $u_\mu$,
and we have reintroduced $\hbar$ and $c$.

In the same way, the pressure on the plates is given by
\begin{equation}
P(z=0)={\pi^{2}\over 240}\frac{\hbar c}{a^{4}}
\left(1+{2\over 3}{ga \over c^{2}}\right), \;
P(z=a)=-{\pi^{2}\over 240 }\frac{\hbar c}{a^{4}}\left(1-{2\over 3}
{ga \over c^{2}}\right),
\label{(5.3)}
\end{equation}
so that a net force pointing upwards along the $z$-axis is obtained,
in full agreement with Eq. (8) in Ref. \cite{Call02}, with magnitude
\be
F={\pi^{2}\over 720}{A \hbar g\over c a^{3}}.
\label{(5.4)}
\ee
The reader may wonder whether the pressure on the outer faces of the 
cavity may alter this result. A simple way to answer this question is to
imagine that our cavity is included into a surrounding cavity on both
sides. On denoting by $b$ the common separation between either
plates of the original cavity and the nearest plate of the surrounding 
cavity, and assuming that $b$ is such that 
$a/b <<1$, but still small enough so as to obtain
$gb/c^{2} << 1$, we see from Sec. IV that the outer pressure on both plates
of the original cavity includes the same divergent contribution which acts
from within plus a finite contribution that becomes negligible for
$a/b <<1$. To sum up, the divergent contributions to the pressure 
from the inside and the outside of either plate cancel each other exactly, 
and one is left just with the finite contribution from the inside,
as given in Eq. (5.3).  

As a check of the result, it can be verified   that the
energy-momentum tensor is covariantly conserved to first order in
$\epsilon$. To this order, the covariant conservation law implies
the following conditions 
\beqa & \epsilon^0: & \langle
T^{(0)\mu\nu}\rangle _{,\nu}=0,
\label{(5.5)} \\
& \epsilon: & \left\{\begin{array}{c}
  \langle T^{(1)ij} \rangle_{,j}=0~~(i=0,1,2), \\
  \\
\frac{1}{2}\left(  \langle T^{(0)00}\rangle + \langle
T^{(0)33}\rangle \right)+ \langle T^{(1)33}\rangle_{,3} =0,
\end{array}\right.
\label{(5.6)} 
\eeqa 
that are indeed satisfied (cf. \cite{Dewi84,Endo84}).

\section{Concluding remarks}

The Casimir effect for scalar fields in curved spacetime
\cite{Borm96} has been previously considered by  various authors,
in a number of different geometries \cite{Bayi93a}--\cite{Saha06}.
 More precisely, Refs. \cite{Bayi93a,
Bayi93b} focus on a massless scalar field in half of the Einstein
static universe, while Ref. \cite{Bayi94} studies the same field
in a  Friedmann background geometry with spherical boundary, and
the associated nonintegrable divergence in the renormalized energy
density. Massless scalar fields in spherical or cylindrical shells
are studied also in Refs. \cite{Seta01}--\cite{Seta05}, with local
boundary conditions of the Dirichlet or Robin type. Moreover, the
work in Ref. \cite{Saha06} deals with the more complicated problem
of a massive nonminimally coupled scalar field in between two
infinite parallel plates moving by uniform proper acceleration.
Such a scalar field is taken to obey Robin boundary conditions on
the plates, and the interaction forces between the plates are
investigated as functions of the proper accelerations and
coefficients in the boundary conditions. Interestingly, for some
values of these parameters the interaction forces are found to be
repulsive at small distances and attractive at large distances
\cite{Saha06}.

To the best of our knowledge, the analysis presented in this paper
represents the first study of the energy-momentum tensor for the
electromagnetic field in a Casimir cavity placed in a weak
gravitational field. The resulting calculations are considerably
harder than in the case of scalar fields. By using Green-function
techniques, we have evaluated the influence of the gravity
acceleration on the regularized energy-momentum tensor of the
quantized electromagnetic field between two plane-parallel ideal
metallic plates, at rest in the gravitational field of the earth,
and lying in a horizontal plane. In particular, we have obtained a
detailed derivation of the theoretical prediction according to
which a Casimir device in a weak gravitational field will
experience a tiny push in the upwards direction \cite{Call02}.
This result is consistent with the picture that the {\it negative}
Casimir energy in a gravitational field will behave like a {\it
negative mass}. Our original results are
relevant both for quantum field theory in curved space-time, and
for the theoretical investigation of vacuum energy effects (see
below).

We stress that in our computation we do not add by hand a mass
term for photons, unlike the work in Ref. \cite{Chri78}, since
this regularization procedure breaks gauge invariance even prior
to adding a gauge-fixing term, and is therefore neither
fundamental nor desirable \cite{Bimo04, Endo84}. In agreement with
the findings of Deutsch and Candelas for conformally invariant
fields \cite{Deut79}, we find that on approaching either wall, the
energy density of the electromagnetic field diverges as the third
inverse power of the distance from the wall. It is interesting to
point out that, in the intermediate stages of the computation,
quartic divergences appear in the contributions from the ghost and
the gauge breaking terms, which however cancel each other exactly.
The occurrence of these higher divergences in such terms is also
consistent with the results of Deutsch and Candelas, in view of
the obvious fact that ghost fields are not ruled by conformally
invariant operators. 

Our results, jointly with the work in Refs. \cite{Call02, Cald02},
are part of a research program aimed at studying the Casimir
effect in a weak gravitational field, with possible corrections
(albeit small) to the attractive force on the plates resulting
from spacetime curvature \cite{Sorg05} (cf. the recent theoretical
analysis of quantum vacuum engineering propulsion in Ref.
\cite{Pint06}). Hopefully, these efforts
represent a first step towards an experimental verification of the
validity of the Equivalence Principle for virtual photons.

\acknowledgments The work of G. Bimonte and G. Esposito has been
partially supported by PRIN {\it SINTESI}. The work of L. Rosa has
been partially supported by PRIN {\it FISICA ASTROPARTICELLARE}.

\appendix
\section{On Ward identities and boundary conditions for gauge fields}

In the case of gauge theories there exists, on the space $\Phi$ of field
histories, a set of vector fields $Q_{\alpha}$ that leave the action
$S$ invariant, i.e. \cite{Dewi84}
\begin{equation}
Q_{\alpha}S=0.
\label{(A1)}
\end{equation}
On denoting by $Q_{\; \alpha}^{i}$ the components of such vector fields,
and by $S_{,i}$ the functional derivatives of the action with respect to
field variables $\varphi^{i}$, Eq. (A1) implies that
\begin{equation}
S_{,i}Q_{\; \alpha}^{i}=0.
\label{(A2)}
\end{equation}
By virtue of Eq. (A2), the operator $S_{,ij}$ is not invertible, and an
invertible operator $U_{ij}$ is obtained upon adding to $S$ the
gauge-breaking term
\begin{equation}
{1\over 2}P^{\alpha}\omega_{\alpha \beta}P^{\beta}
={1\over 2}\int d^{4}x \int d^{4}x' P^{\alpha}(x)
\omega_{\alpha \beta}(x,x') P^{\beta}(x'),
\label{(A3)}
\end{equation}
where $P^{\alpha}$ is the gauge-fixing functional and
$\omega_{\alpha \beta}(x,x')=\omega_{\alpha \beta'}$ is a
nonsingular, symmetric continuous matrix, possibly depending on
field variables to achieve full covariance of the formalism
\cite{Dewi81}. The desired invertible gauge-field operator
$U_{ij}$ reads therefore
\begin{equation}
U_{ij} \equiv S_{,ij}+P_{\; ,i}^{\alpha} \; \omega_{\alpha \beta}
\; P_{\; ,j}^{\beta},
\label{(A4)}
\end{equation}
which should be considered jointly with the ghost operator
\begin{equation}
{\widehat U}_{\; \beta}^{\alpha} \equiv Q_{\beta}P^{\alpha}
=P_{\; ,i}^{\alpha} \; Q_{\; \beta}^{i},
\label{(A5)}
\end{equation}
since the full action in the functional integral for the
$\langle {\rm out} | {\rm in} \rangle$ amplitude is given by
$$
S+{1\over 2}P^{\alpha}\omega_{\alpha \beta}P^{\beta}
+\chi_{\alpha}{\widehat U}_{\; \beta}^{\alpha}\psi^{\beta},
$$
with $\chi_{\alpha}$ and $\psi^{\beta}$ independent ghost fields
\cite{Dewi84}, which obey Fermi statistics if the gauge fields
$\varphi^{i}$ are bosonic.

Repeated functional differentiation of Eq. (A2) yields the classical
Ward identities of the theory, which can be used to derive remarkable
identities among the gauge-field and ghost Green functions. For example,
the first functional derivative of Eq. (A2) yields
\begin{equation}
S_{,ij}Q_{\; \alpha}^{i}+S_{,i}Q_{\; \alpha ,j}^{i}=0.
\label{(A6)}
\end{equation}
We now re-express $S_{,ij}$ from Eq. (A4) and bear in mind the definition
(A5) to find
\begin{equation}
U_{ij}Q_{\; \alpha}^{j}-{\widehat U}_{\; \alpha}^{\beta}
\; \omega_{\beta \delta} \; P_{\; ,j}^{\delta}
+S_{,i}Q_{\; \alpha ,j}^{i}=0.
\label{(A7)}
\end{equation}
Restriction to the dynamical subspace, where $S_{,i}$ vanishes, and
composition with the ghost Green function
${\widehat G}^{\alpha \gamma}$, for which
\begin{equation}
{\widehat U}_{\; \alpha}^{\beta} \;
{\widehat G}^{\alpha \gamma}=-\delta^{\beta \gamma},
\label{(A8)}
\end{equation}
yields
\begin{equation}
U_{ij}Q_{\; \alpha}^{i} \; {\widehat G}^{\alpha \gamma}
+\omega_{\; \delta}^{\gamma} \; P_{\; ,j}^{\delta}=0.
\label{(A9)}
\end{equation}
Now we act on Eq. (A9) with the gauge-field Green function $G^{jk}$
(see Eq. (3.1)), finding therefore \cite{Dewi81}
\begin{equation}
Q_{\; \alpha}^{k} \; {\widehat G}^{\alpha \gamma}
=\omega_{\; \delta}^{\gamma} \; P_{\; ,j}^{\delta} \; G^{jk}.
\label{(A10)}
\end{equation}
This equation holds for all type-I theories, i.e. all gauge theories
for which the Lie bracket of the vector fields in Eq. (A1) is a linear
combination of such fields with structure constants, i.e.
\begin{equation}
[Q_{\alpha},Q_{\beta}]=C_{\; \alpha \beta}^{\gamma} \; Q_{\gamma},
\label{(A11)}
\end{equation}
with $C_{\; \alpha \beta ,i}^{\gamma}=0$. For the case of Maxwell
theory in curved spacetime, the ghost Green function has no
explicit group indices, $Q_{\; \alpha}^{k}$ reduces to covariant
derivatives with respect to the Levi--Civita connection, and
$\omega_{\; \delta}^{\gamma} \; P_{\; ,j}^{\delta} G^{j k}$ yields
the covariant derivative of the photon Green function. Thus, the
Ward identity (3.33) is eventually obtained. For the sake of
completness we write down the expansion to first order in
$\epsilon$ of the first of Eqs. (3.33): 
\beqa &\epsilon^0)&
\eta^{\mu\nu} H^{(0)}_{\mu\alpha',\nu}+H^{(0)}_{,\alpha'}=0
~~\Rightarrow ~~ \left\{\begin{array}{c}
  -H^{(0)}_{00',0}+H^{(0)}_{,0'}=0, \\
 H^{(0)}_{ii',i}+H^{(0)}_{,i'}=0,
\label{(A12)}
\end{array} \right. \\
&\epsilon)&
\eta^{\mu\nu}H^{(1)}_{\mu\alpha',\nu}
-a \eta^{\mu\nu}\frac{\Gamma^\rho_{\mu\nu}}{\epsilon}
H^{(0)}_{\rho\alpha'}
+z\delta^\mu_0\delta^\nu_0 H^{(0)}_{\mu\alpha',\nu}
+H^{(1)}_{,\alpha'}=0~~\Rightarrow \nonumber\\
& & \left\{\begin{array}{c}
  -H^{(1)}_{00',0} -H^{(1)}_{30',3}
+ zH^{(0)}_{00',0}+H^{(1)}_{,0'}=0, \\
 \\
 H^{(1)}_{ii',i}+H^{(0)}_{,i'}=0~~(i=1,2),\\
 \\
   -H^{(1)}_{03',0} -H^{(1)}_{33',3}
+ \frac{1}{2}H^{(0)}_{33'}+H^{(1)}_{,3'}=0.
\end{array} \right .
\label{(A13)}
 \eeqa

A recipe for writing gauge-invariant boundary conditions
\cite{Avra99} in field theory is expressed by the equations
\begin{equation}
\Bigr[\pi_{\; j}^{i} \; \varphi^{j}\Bigr]_{\partial M}=0,
\label{(A14)}
\end{equation}
\begin{equation}
\Bigr[P^{\alpha}[\varphi]\Bigr]_{\partial M}=0,
\label{(A15)}
\end{equation}
\begin{equation}
[\psi_{\beta}]_{\partial M}=0,
\label{(A16)}
\end{equation}
where $\pi_{\; j}^{i}$ is a tangential projection operator. In the case
of Maxwell theory, considered in our paper, Eqs. (A14)--(A16) become
\begin{equation}
[A_{k}]_{\partial M}=0,
\label{(A17)}
\end{equation}
\begin{equation}
[\Phi(A)]_{\partial M}=0,
\label{(A18)}
\end{equation}
\begin{equation}
[\lambda]_{\partial M}=0.
\label{(A19)}
\end{equation}
On performing the familiar gauge transformation
\begin{equation}
{ }^{\lambda}A_{\mu} \equiv A_{\mu}+\nabla_{\mu}\lambda,
\label{(A20)}
\end{equation}
Eq. (A17) is preserved under (A20) if and only if $\lambda$ vanishes
on the boundary (see (A19)), since tangential derivatives and restriction
to the boundary are commuting operations \cite{Avra99}. Moreover, under
(A20), the gauge-fixing functional changes according to
\begin{equation}
\Phi(A)-\Phi({ }^{\lambda}A)={\widehat U} \; \lambda,
\label{(A21)}
\end{equation}
where ${\widehat U}$ is a linear differential operator. Now if $\lambda$
is expanded according to a complete orthonormal set of eigenfunctions of
${\widehat U}$, i.e.
\begin{equation}
{\widehat U} \; u_{i}=\mu_{i} \; u_{i},
\label{(A22)}
\end{equation}
\begin{equation}
\lambda=\sum_{i}C_{i} \; u_{i},
\label{(A23)}
\end{equation}
the gauge invariance of the boundary condition (A18) is again guaranteed
by (A19), because
\begin{equation}
\Phi(A)-\Phi({ }^{\lambda}A)={\widehat U} \; \lambda
=\sum_{i}C_{i} \; \mu_{i} \; u_{i}.
\label{(A24)}
\end{equation}
The vanishing of $u_{i}$ on the boundary implies therefore that both
$\lambda$ and $\Phi({ }^{\lambda}A)$ vanish therein, if $\Phi(A)$ was
already satisfying Eq. (A18). In the full quantum theory, $\lambda$
should be replaced by two fermionic ghost fields \cite{Dewi84, Espo97}.

\section{Green functions and energy-momentum tensors}

\subsection{The Fourier transform of $G^{(1)}_{\lambda\nu'}$ }

In the following, for the sake of simplicity, we always assume
$z'\geq z$, hence we have to be careful when computing integrals and
limits, but the resulting formulae become relatively less
cumbersome. With the notation in Eq. (3.32) we find
(hereafter for the sake of brevity we define
$\xi\equiv a\kappa,~s(\xi)\equiv z \kappa=\frac{ z \xi}{a}$
and $s'(\xi)\equiv z' \kappa=\frac{ z' \xi}{a}$)
\begin{eqnarray}
\Phi^{00'} & = & \frac{a \sin^{-2}(\xi)}{8 \xi^4}
\Biggl[-a^2 \omega^2 \cos (2 \xi-s-s') s^2+a^2 \omega^2 \cos (s) \cos
   (s') s^{2}\nonumber\\
 &+&  \sin (s) \Biggl(a^2 \omega^2 \left((s^2-s'^2) \sin (2
   \xi-s')+\left(2 \xi^2-s'^2\right) \sin (s')\right) \nonumber\\
 &-&  2 \left(\xi^2+a^2
   \omega^2\right) (s+s') \sin (\xi) \sin (\xi-s')\Biggr)\Biggr], \\
\Phi^{03'}&= &-\frac{i a^2}{2 \xi^3}\omega \sin^{-1}(\xi)
\sin (s) ((s-s') \cos
   (\xi-s')-\sin (\xi-s')), \\
\Phi^{11'}&= &   \frac{a \sin^{-2}(\xi) }{8 \xi^4}
\Biggl[a^2 \omega^2 \cos (2 \xi-s-s') s^2-a^2 \omega^2 \cos (s) \cos (s')
   s^{2} \nonumber \\
&+&   \sin (s) \Biggl(a^2 \omega^2
\left(\left(s'^2-s^2\right) \sin (2 \xi-s')+\left(s'^2-2
   \xi^2\right) \sin (s')\right) \nonumber \\
&-&   2 (\xi^2-a^2 \omega^2) (s+s') \sin (\xi) \sin
   (\xi-s')\Biggr)\Biggr], \\
\Phi^{22'}&= &\Phi^{11'}, \\
\Phi^{30'}&= &-\frac{i a^2 \omega \sin^{-1}(\xi) }
{2 \xi^3}\biggl((s'-s) \cos (s)+\sin (s)\biggr) \sin (\xi-s'), \\
\Phi^{33'}&= &\frac{a \sin^{-1}(\xi) }{8 \xi^4}
\Biggl[2 \left(\xi^2-a^2  {\omega}^2
   \left(s^2-1\right)\right) \cos (\xi- {s'}) \sin (s) \nonumber \\
 &+&  \cos (s) \Biggl(\csc
   (\xi) \left(\left(\xi^2-a^2  {\omega}^2 \left( {s'}^2-1\right)\right)
   \cos (2 \xi- {s'}) \right. \nonumber\\
&+& \left.  \cos ( {s'}) \left(-\left(2 a^2
    {\omega}^2+1\right) \xi^2+a^2  {\omega}^2
   \left( {s'}^2-1\right)+\left(\xi^2-a^2  {\omega}^2\right) (s+ {s'})
   \sin (2 \xi)\right)\right) \nonumber\\
&+&   2 \left(\xi^2-a^2  {\omega}^2\right) (s+ {s'})
   \sin (\xi) \sin ( {s'})\Biggr)\Biggr].
\end{eqnarray}

\subsection{The energy-momentum tensors}

In our analysis we deal with
\be
\langle T^{(0)\mu\nu'}\rangle = 2 i
\int{ \frac{d\omega d^2k }{(2\pi)^3} e^{-i\omega(t-t')+ i
{\vec k}_\perp\cdot({\vec x}-{\vec x}')_\perp} }
\widetilde{T}^{(0)\mu\nu'}[\omega,\vec{k};z,z'],
\label{(B7)}
\ee
where the $2i$ factor results from the relation (3.3) between the
Hadamard and Feynman Green functions, and
$\widetilde{T}^{(0)\mu\nu'}$ is a symmetric tensor whose components are
\begin{eqnarray}
\widetilde{T}^{(0)00'} &=& -\frac{ a\,{\omega}^2}{2\xi}
\,\cos (\xi + s - s')\,\csc (\xi),   \nonumber \\
\widetilde{T}^{(0)01'} &=& -\frac{ a\,k_x\,\omega}{2\xi}\,
      \cos (\xi + s - s')\,\csc (\xi),  \nonumber \\
\widetilde{T}^{(0)02'} &=& -\frac{ a\,k_y\,\omega}{2\xi}\,
      \cos (\xi+ s - s')\,\csc (\xi), \nonumber \\
\widetilde{T}^{(0)03'} &=& -\frac{i }{2}\,\omega\,\csc (\xi)\,
  \sin (\xi + s - s'), \nonumber \\
\widetilde{T}^{(0)11'} &=& \frac{\xi^2 + a^2\,
       \left( k_y^2 - \omega^2
         \right) }{2\,a\,\xi}\,
    \cos (\xi + s - s')\,\csc (\xi), \nonumber \\
\widetilde{T}^{(0)12'} &=& -\frac{ a\,k_x\,k_y}{2\xi}\,
      \cos (\xi + s - s')\,\csc (\xi), \nonumber \\
\widetilde{T}^{(0)13'} &=& -\frac{i }{2}\,k_x\,
  \sin (\xi + s - s')\,\csc (\xi), \nonumber \\
\widetilde{T}^{(0)22'} &=& -\frac{a\,{k_y}^2 }{2\xi}\,
      \cos (\xi + s - s')\,\csc (\xi), \nonumber \\
\widetilde{T}^{(0)23'} &=& -\frac{i }{2}\,k_y\,\,
  \sin (\xi + s - s')\csc (\xi), \nonumber \\
\widetilde{T}^{(0)33'} &=& -\frac{\xi}{2\,a}\,
\cos (\xi+ s - s')\,\csc (\xi). \nonumber
\end{eqnarray}
Now we first take the limit as $t' \rightarrow t$ and
${\vec x}_{\perp}' \rightarrow {\vec x}_{\perp}$. At that stage,
on taking $z\neq z'$ we pass the limit under the integral, then
sending $\omega\rightarrow i\omega,~(\kappa\rightarrow
{i \kappa})$, and going to spherical
coordinates: ${\omega}^{2} \rightarrow -\xi^{2} \; \cos^{2}\theta$,
$k_{y}^{2} \rightarrow
\xi^{2} \; \sin^{2}\theta \; \sin^{2}\phi$, angular integration
yields eventually
\be
\langle T^{(0)\mu\nu'} \rangle[z,z']
=\frac{1}{6\,{\pi }^2} \frac{1}{a^4}\int d\xi \; \xi^{3} \times
\,
\cosh (\xi-s-s')\,{\mbox{csch}}(\xi)
{\rm diag}(-1,1,1,-3).
\label{(B8)}
\ee
After integrating over the $\xi$ variables we
obtain Eq. (\ref{(4.1)}). Analogously, upon integrating
over the solid angle we get, to first order in momentum space,
\begin{eqnarray}
\langle T^{(1)00} \rangle [z,z'] &=& \frac{\pi}{60 a^{3}} \int
d\xi \;\xi^2 {\sinh^{-2}}(\xi) \,
    \biggr \{ -3(2+s^{2})\xi^{2}\cosh[\xi(s-s')]
-8 \cosh[\xi(-2+s+s')] \nonumber \\
&+& 8 \cosh[\xi(s+s')] \nonumber \\
&+& s \xi \Bigr[3s \xi \cosh[\xi(2+s-s')]
+32 \Bigr(\sinh[\xi(2+s-s')]-\sinh[\xi(s-s')]\Bigr)\Bigr]
\nonumber \\
&+& 3\xi s' \Bigr[-4 \sinh[\xi(s-s')]
+4 \sinh [\xi(2+s-s')] \nonumber \\
&+& \xi s' \Bigr(\cosh[\xi(s-s')]
-\cosh[\xi(2+s-s')] \Bigr) \Bigr] \biggr \}. \\
\langle T^{(1)11} \rangle [z,z']&=& \frac{1}{120 a^{3}\pi^{2}}\int
d\xi \xi^2   \, {\sinh^{-1}}(\xi)\,
    \Biggr[ -4\,\left( {s} + s' \right) \,\cosh (\xi + s - s') +
      \biggl( -2\,\cosh (2\,\xi - s - s')  \nonumber \\
&+&  {{\xi}}^2\,
          \cosh (s - s') +2\,\cosh (s + s') \biggr) \,
       {\sinh^{-1}}(\xi) \nonumber \\
&+&
      \left( -{{s}}^2 +
         {{s'}}^2 \right) \,
       \sinh (\xi + s - s') \Biggr], \\
\langle T^{(1)22} \rangle [z,z'] &=&T^{(1)11'}, \\
\langle T^{(1)33} \rangle [z,z'] &=& \frac{1}{24 a^{3} \pi^{2}} \int
d\xi  \xi^2  \, {\sinh^{-1}}(\xi)\,
    \Biggl[ -\left( {{\xi}}^2\,
         \cosh (s - s')\,
         {\sinh^{-1}}(\xi) \right)  \nonumber \\
&+&   \left( {s} + s' \right)
         \,\biggl( 4\,\cosh (\xi + s - s') + \left( {s} - s'
            \right) \,\sinh (\xi + s - s')
         \biggr)  \Biggr],
\end{eqnarray}
where we have written explicitly only nonvanishing components.
Integration over $\xi$ finally yields Eqs. (4.4)--(4.7).

For completeness we write also the ghost energy-momentum tensor
$\langle T^{\mu\nu}\rangle_{{\rm gh}}$
for Dirichlet boundary conditions, i.e.
\beqa
\langle T^{(0)00} \rangle_{\rm gh}
&=& \frac{{\pi}^2}{720\,a^4}
   + \frac{{\pi }^2\,\left( 2 +
       \cos (\frac{2\,\pi \,z}{a}) \right) \,
     {\csc^{4} (\frac{\pi \,z}{a})}}{24\,a^4}+
  \lim_{z'->z^+} \frac{1}{{\pi }^2\, (z-z')^{4}},
\label{(B13)} \\
 \langle T^{(0)11} \rangle_{\rm gh} &=& -\frac{{\pi }^2}{720\,a^4}
   - \frac{{\pi }^2\,\left( 2 +
       \cos (\frac{2\,\pi \,z}{a}) \right) \,
     {\csc^{4} (\frac{\pi \,z}{a})}}{24\,a^4}-
   \lim_{z'->z^+}\frac{1}{{\pi }^2\, (z-z')^{4}},
\label{(B14)} \\
\langle T^{(0)22} \rangle_{\rm gh} &=& \langle T^{(0)11} \rangle_{\rm gh},
\label{(B15)} \\
 \langle T^{(0)33} \rangle_{\rm gh}
&=& \frac{\pi^2}{240 a^4}+\lim_{z'->z^+}\frac{3}{\pi^2(z-z')^4},
\label{(B16)}
\eeqa
and
\beqa
\langle T^{(1)00} \rangle_{\rm gh} &=& \frac{{\pi }^2\,\left(
3\,a - 11\,z \right)}{3600\,a^4}+\frac{{\pi }^2}
{40\,a^5}\,\left(  a\,\left( a - 7\,z \right)
-\pi z(a-z)\cot(\frac{\pi\,z}{a}) \right) \,{\csc^{4} (\frac{\pi \,z}{a})}
\nonumber \\
&+& \frac{\pi }{240\,a^5}\,\csc^2(\frac{\pi\,z}{a})
\left( (5\,a^2+2\pi^2z(a-z))\,{\cot(\frac{\pi \,z}{a})}-4a\pi(a-7z) \right)
\nonumber \\
&-&\lim_{z'->z^+}  \frac{z}{{\pi }^2\, (z-z')^{4}},
\label{(B17)} \\
\langle T^{(1)11} \rangle_{\rm gh}
&=&-\frac{\left( {\pi }^2\,\left( a - 2\,z \right)  \right)}{3600\,a^4}
 \nonumber \\
&-& \frac{{\pi }^2}{160\,a^6}\left( a( 8a^2 -15a z-9z^2)
-8\pi z(a^2-az-z^2)\cot(\frac{\pi \,z}{a})\right)
\,{\csc^{4} (\frac{\pi \,z}{a})} \nonumber \\
&-&  \frac{\pi}{480\,a^6}
 \,\Biggl( (10a^3+a^2(3+8\pi^2)z -8\,\pi^2z^2(a+z))\cot(\frac{\pi\,z}{a})
\nonumber \\
 &+&   2 a \pi (-8 a^2+15a z+9z^2)\Biggr) {\csc^{2} (\frac{\pi \,z}{a})},   \,
\label{(B18)}   \\
  \langle T^{(1)22}  \rangle_{\rm gh}
&=& \langle T^{(1)11} \rangle_{\rm gh},
\label{(B19)}\\
 \langle T^{(1)33} \rangle_{\rm gh} &=&\frac{{\pi }^2(a-2z)}{720\,a^4}
 - \frac{\pi{\csc^{2} (\frac{\pi \,z}{a})}}{96\,a^4}
  \cot (\frac{\pi \,z}{a}) \nonumber \\
&-&  \frac{1 }{64\pi^2\,a^4}\,
\biggr(4a\Bigr(\psi^{(1)}(1/2-z/a) -\psi^{(1)}(-1/2+z/a)\Bigr)  \nonumber \\
&+&  (a-2z)\Bigr(\psi^{(2)}(1/2-z/a)+ \psi^{(2)}(-1/2+z/a)\Bigr)\biggr),
\label{(B20)}
\eeqa
where $\psi^{(n)}(z)$ is the $n$-th derivative of the
logarithmic derivative $\psi(z)$ of the $\Gamma$-function.


\begin{references}
\bibitem{Chri76}
S.M. Christensen, Phys. Rev. D {\bf 14}, 2490 (1976).
\bibitem{Chri78}
S.M. Christensen, Phys. Rev. D {\bf 17}, 946 (1978).
\bibitem{Bunc79}
T.S. Bunch and L. Parker, Phys. Rev. D {\bf 20}, 2499 (1979).
\bibitem{Deut79}
D. Deutsch and P. Candelas, Phys. Rev. D {\bf 20}, 3063 (1979).
\bibitem{Brow86}
M.R. Brown and A.C. Ottewill, Phys. Rev. D {\bf 34}, 1776 (1986).
\bibitem{Bimo04}
G. Bimonte, E. Calloni, L. Di Fiore, G. Esposito, L. Milano, and
L. Rosa, Class. Quant. Grav. {\bf 21}, 647 (2004).
\bibitem{Bord01}
M. Bordag, U. Mohideen, and V.M. Mostepanenko, Phys. Rep. {\bf 353},
1 (2001).
\bibitem{Bart01}
G. Barton, J. Phys. A {\bf 34}, 4083 (2001).
\bibitem{Kamp68}
N.G. van Kampen, B.R. Nijboer, and K. Schram, Phys. Lett. A {\bf 26},
307 (1968).
\bibitem{Grah02}
N. Graham, R.L. Jaffe, V. Khemani, M. Quandt, M. Scandurra, and H.
Weigel, Nucl. Phys. B {\bf 645}, 49 (2002).
\bibitem{Davi05}
P.C.W. Davies, J. Opt. B {\bf 7}, S40 (2005); M. Ishak,
astro-ph/0504416.
\bibitem{Call02}
E. Calloni, L. Di Fiore, G. Esposito, L. Milano, and L. Rosa,
Phys. Lett. A {\bf 297}, 328 (2002).
\bibitem{Misn73}
C. Misner, K.P. Thorne, and J.A. Wheeler, {\it Gravitation}
(Freeman, S. Francisco, 1973).
\bibitem{Marz94}
K.P. Marzlin, Phys. Rev. D {\bf 50}, 888 (1994).
\bibitem{Dewi75}
B.S. DeWitt, Phys. Rep. {\bf 19C}, 295 (1975).
\bibitem{Lore67}
L. Lorenz, Phil. Mag. {\bf 34}, 287 (1867).
\bibitem{Dewi84}
B.S. DeWitt, `The Spacetime Approach to Quantum Field Theory', in
{\it Relativity, Groups and Topology II}, eds. B.S. DeWitt and
R. Stora (North--Holland, Amsterdam, 1984).
\bibitem{RuSy31}
H.S. Ruse, Proc. London Math. Soc. {\bf 32}, 87 (1931);
J.L. Synge, Proc. London Math. Soc. {\bf 32}, 241 (1931);
J.L. Synge, {\it Relativity: The General Theory}
(North--Holland, Amsterdam, 1960).
\bibitem{Espo97}
G. Esposito, A.Yu. Kamenshchik, and G. Pollifrone, {\it Euclidean
Quantum Gravity on Manifolds with Boundary}, Fundamental
Theories of Physics, Vol. {\bf 85} (Kluwer, Dordrecht, 1997).
\bibitem{Endo84}
R. Endo, Prog. Theor. Phys. {\bf 71}, 1366 (1984).
\bibitem{Borm96}
K. Bormann and F. Antonsen, `The Casimir Effect of Curved Space-Time
(formal developments)', in {\it Proceedings of the 3rd International
A. Friedmann Seminar} (Friedmann Lab. Press, St. Petersburg, 1995).
[hep-th/9608142]
\bibitem{Bayi93a}
S.S. Bayin and M. \"{O}zcan, Class. Quantum Grav. {\bf 10}, L115 (1993).
\bibitem{Bayi93b}
S.S. Bayin and M. \"{O}zcan, Phys. Rev. D {\bf 48}, 2806 (1993).
\bibitem{Bayi94}
S.S. Bayin and M. \"{O}zcan, Phys. Rev. D {\bf 49}, 5313 (1994).
\bibitem{Seta01}
M.R. Setare and R. Mansouri, Class. Quant. Grav. {\bf 18}, 2331 (2001).
\bibitem{Seta03}
M.R. Setare, hep-th/0308108.
\bibitem{Seta04a}
M.R. Setare and M.B. Altaie, Gen. Rel. Grav. {\bf 36}, 331 (2004).
\bibitem{Seta04b}
M.R. Setare, Gen. Rel. Grav. {\bf 36}, 1965 (2004).
\bibitem{Seta05}
M.R. Setare and F. Darabi, hep-th/0511077.
\bibitem{Saha06}
A.A. Saharian, R.M. Avagyan, and R.S. Davtyan, Int. J. Mod. Phys. A
{\bf 21}, 2353 (2006).
\bibitem{Cald02}
R.R. Caldwell, `Gravitation of the Casimir effect and the cosmological
nonconstant' (astro-ph/0209312).
\bibitem{Sorg05}
F. Sorge, Class. Quant. Grav. {\bf 22}, 5109 (2005).
\bibitem{Pint06}
F. Pinto, J. Brit. Interpl. Soc. {\bf 59}, 247 (2006).
\bibitem{Dewi81}
B.S. DeWitt, `A Gauge-Invariant Effective Action', in {\it Quantum
Gravity 2, A Second Oxford Symposium}, eds. C.J. Isham, R. Penrose, and
D.W. Sciama (Clarendon Press, Oxford, 1981).
\bibitem{Avra99}
I.G. Avramidi and G. Esposito, Commun. Math. Phys. {\bf 200}, 495 (1999).
\end{references}
\end{document}